\documentclass[preprint]{revtex4}
\usepackage{graphicx}
\usepackage{epsfig}
\input{epsf}

\begin{document}

\title{Quantitative Analysis of Hydrogenated DLC Films by Visible Raman Spectroscopy}
\author{Achintya Singha$^1$, Aditi Ghosh$^1$, Nihar Ranjan
Ray$^{2}$}\email{niharranjan.ray@saha.ac.in}

\author{Anushree Roy$^{1}$} \email{anushree@phy.iitkgp.ernet.in}
\affiliation{$^1$Department of Physics,Indian Institute of Technology,Kharagpur 721302, India\\
$^2$Plasma Physics Division, Saha Institute of Nuclear Physics,
Kolkata 700064, India}

\begin{abstract}
The correlations between properties of hydrogenated diamond like
carbon films and their Raman spectra have been investigated. The
films are prepared by plasma deposition technique, keeping
different hydrogen to methane ratio during the growth process. The
hydrogen concentration, sp$^3$ content, hardness and optical Tauc
gap of the materials have been estimated from a detail analysis of
their Raman spectra. We have also measured the same parameters of
the films by using other commonly used techniques, like sp$^3$
content in films by x-ray photoelectron spectroscopy, their Tauc
gap by ellipsometric measurements and hardness by micro-hardness
testing. The reasons for the mismatch between the characteristics
of the films, as obtained by Raman measurements and by the above
mentioned techniques, have been discussed. We emphasize on the
importance of the visible Raman spectroscopy in reliably
predicting the above key properties of DLC films.

\noindent Keywords : Diamond like carbon, Raman scattering.
\end{abstract}

 \maketitle

\section{introduction}
In recent years, hard diamond like carbon (DLC) films have
attracted a great deal of research interest as they are useful
materials for protective optical and tribological coating
\cite{geis:1993,koidl:1989,angus:1988,robertson:1992,tsai:1987}.
These films can be of greater economic importance in the long run,
if their properties can be understood to the extent we know the
bulk characteristics of diamond today. The characteristics of DLC
films depend considerably on the atomic structure of the films.
For example, opaque samples with hardness one-fifth of that for
diamond and the transparent ones nearly as dense and hard as
diamond \cite{tamor:1994}, have been considered as DLC films, for
which specific properties have been investigated. This indicates
that over a wide range of atomic structures one can achieve
diamond-like behavior.

DLC films can be defined as composites of nanocrystalline diamond
and/or amorphous carbon with/without hydrogen (required to
passivate the dangling bonds of carbon). The  films with 20 - 50
$\%$ of hydrogen content is commonly known as hydrogenated DLC
(HDLC) films. DLC films are usually composed of amorphous carbon
with a significant amount of sp$^3$ bonds \cite{ferrari:2000}. The
films with higher (more than 85\%) sp$^3$ content are known as
tetrahedral amorphous carbon (ta-C) rather than DLC. Such films
along with a considerable amount of hydrogen content are known as
ta-C:H. In general, the amorphous carbon can have sp$^3$
(diamond-like), sp$^2$ (graphite-like) and even sp$^1$ sites. The
properties of any carbon material depend on the ratio of the
amount of sp$^2$ to that of the sp$^3$ bonds. In addition to the
sp$^3$ content, the clustering of sp$^2$ phase also plays an
important role in determining different properties of these
materials, especially, their optical, electrical and mechanical
characteristics. Moreover, the presence of nanocrystalline (NC)
diamond in DLC/HDLC films results in a special characteristic of
these materials compared to graphite or diamond. It has been shown
by Badziag\emph{ et al.} that the nanodiamond clusters of less
than 5 nm in diameter are more stable than the graphite clusters
of same size \cite{badziag:1990}.

A variety of analytic techniques have been used to characterize
the DLC/HDLC films. Common tools to measure sp$^2$ and sp$^3$
contents in the carbon films are nuclear magnetic resonance (NMR)
\cite{kaplan:1985,tamor:1991} or electron energy loss spectroscopy
(EELS) \cite{fink:1983,fallon:1993}.  X-ray photoelectron (XP)
spectroscopy has also been used to estimate sp$^3$ content in DLC
films \cite{yan:2004}. Ellipsometric measurement is in general
used to measure the Tauc gap of the material
\cite{franta:2002,vinnichenko:2004}. The standard techniques to
measure hardness of DLC films are nano-indentation or micro-Vicker
hardness testing \cite{marchon:1997, wazumi:2005}. All the above
techniques have their own limitations when used for DLC films. For
example, NMR and EELS are time-consuming and destructive
techniques. The aim of this article is to show that Raman
spectroscopy, a non-destructive probe, can be employed as a single
technique, to predict all the above critical properties of
DLC/HDLC films from a single measurement with better accuracy.

\subsection{Raman modes}

The Raman spectrum of diamond consists of T$_{2g}$ mode at 1332
cm$^{-1}$ (sp$^3$ mode). The phonon confinement in NC diamond
results in a downshift in the Raman spectrum of the diamond line.
The amount of shift depends on the grain size. For grains of less
than 1 nm in size, the maximum of the vibrational density of
states appears at 1260 cm$^{-1}$ \cite{ferrari:2001}. On the other
hand, the Raman spectrum of graphite usually shows two modes, zone
center E$_{2g}$ phonon mode at around 1580-1600 cm$^{-1}$
(commonly known as G-peak, sp$^2$ mode) and K-point phonons at
around 1350 cm$^{-1}$ (commonly known as D-peak : disordered
allowed zone edge mode of graphite). The sp$^1$ mode is present in
the HDLC film in a negligible amount. In HDLC films, other than NC
diamond, the G- and D- peaks, features near 1150 cm$^{-1}$ and its
companion mode 1450 cm$^{-1}$ appear due to sum and difference in
combinations of C=C chain stretching and CH wagging modes [v$_1$
and v$_3$ modes of transpolyacetylene (Trans-PA)] lying in the
grain boundary \cite{ferrari:2001}. Trans-PA is an alternate chain
of sp$^2$ carbon atoms, with a single hydrogen bonded to each
carbon atom. In addition, recent work has shown that a
particularly stable defect in diamond is the dumbbell defect,
which consists of $<$100$>$ split interstitials \cite{prawer:2000}
in HDLC films. This localized defect consists of an isolated
sp$^2$ bonded pair. In vibrational density of states calculation
of a largely four folded amorphous carbon network there is a
strong evidence for paired three-fold coordinated defects which
appear as a sharp localized mode at about 1600 cm$^{-1}$
\cite{drabold:1994}.

\subsection{Three Stage model}

At this point it is to be noted that there is an inherent problem
in applying visible Raman spectroscopy directly to estimate the
ratio of sp$^3$ and sp$^2$ fraction in carbon materials by
measuring the intensities of corresponding Raman modes : the
visible Raman scattering cross-section is 50-230 times more
sensitive to sp$^2$ sites compared to that of sp$^3$ sites, as
visible photons preferentially excite the $\pi$-states of sp$^2$
sites. It was believed that  visible Raman spectroscopy has a
limited use to characterize DLC/HDLC films, especially to estimate
the content of sp$^3$ and sp$^2$ fraction in these materials.
Recently, based on atomic and electronic structure of disordered
carbon, Ferrari and Robertson \cite{ferrari:2000} have proposed a
\emph{three-stage model} and have shown that disordered, amorphous
and diamond like carbon phases in amorphous C-H films can be
characterized by measuring the position and width of G-peak and
intensity ratio of G- and D-peaks in  Raman spectra, rather than
by directly measuring their intensities.  The changes in line
shape of the Raman spectrum for carbon material, when its phase
changes from graphite to NC carbon (stage one) to amorphous carbon
(stage two) and then to ta-C carbon with about 85-90 $\%$ sp$^3$
bonding (stage three), have been explicitly shown in their
article. During the first stage, with an increase in sp$^3$
content in the material, the ratio of the intensity of D peak
(I$_D$) to that of G peak (I$_G$) increases from 0.0 to 2.0 and
simultaneously,  the G peak position ($\omega_{G}$) increases from
1580 cm$^{-1}$ to 1600 cm$^{-1}$. However, in the second stage, a
reverse trend is observed for both parameters with an increase in
sp$^3$ content : the ratio I$_D$/I$_G$ decreases from 2.0 to 0.25,
whereas, the value of $\omega_G$ decreases from 1600 to 1510
cm$^{-1}$. For a phase transition from amorphous carbon to ta-C
phase, $\omega_G$ increases (from 1520 to 1560 cm$^{-1}$) with an
increase in sp$^3$ content and the intensity ratio I$_D$/I$_G$
drops down further (from 0.25 to 0) from what was observed in
stage two. For each stage, relations between G-peak position and
sp$^3$ content as well as Tauc gap of the carbon material have
been discussed by authors.

In this article, we show how visible Raman spectra can be used as
a fingerprint to characterize the DLC/HDLC -films grown by plasma
deposition technique. Our emphasis will be on  four important
characteristics of HDLC films : sp$^3$ content, hydrogen
concentration, optical gap and hardness. Section II covers the
sample preparation technique, which we have followed and other
details regarding the instruments, which we have used for various
measurements. In Section III, we have analyzed our experimental
Raman spectra using  the \emph{three-stage model} proposed by
Ferrari and Robertson, mentioned above. The analysis provides an
insight into the chemical composition (eg. hydrogen content,
sp$^3$ content) and hardness of these films. The estimated sp$^3$
content and hardness of the films, measured from Raman data
analysis, have been compared with the results obtained from the XP
spectroscopic and nano-indentation measurements, respectively. We
could also make an approximate judgement of the optical Tauc gap
in these films from the shift in $\omega_G$ in their Raman
spectra. Later on, the measured gap energies have been compared
with the values we get from ellipsometric measurements. Finally,
in section IV, we have discussed our results with a few concluding
remarks.

Here, we would like to mention that in the article
\cite{tamor:1994}, Tamor and Vassel  described a systematic study
of the Raman spectra of amorphous carbon films. Based on
experimental data, this phenomenological report revealed a clear
correlation between the variation in the G-peak position and the
change in optical gap and hardness of films. In this present
report, based on the theoretical understanding proposed by Ferrari
and Robertson \cite{ferrari:2000}, we have shown how Raman
measurements can be used to study HDLC films in more detail. We
have also shown the presence of NC diamond in our films.
Furthermore, in this report, we have pointed out drawbacks of
other commonly used probes to study this system.

\section{Experiments}
Carbon thin films are deposited on mirror-polished Si(100)
substrate at room temperature using asymmetric capacitively
coupled RF (13.56 MHz) plasma system. The depositions are made
systematically as follows: a pretreatment of the bare mirror
polished Si (100) substrate has been done for 15 minutes using
pure hydrogen plasma at pressure ~0.2 mbar and dc self-bias of
-200 volts. The  deposition has been made for 30 minutes at
pressure ~ 0.7 mbar keeping the flow rate of helium (He) at 1500
sccm, hydrogen (H$_2$) at 500 sccm, and varying the flow rate of
methane (CH$_4$). Five samples (for which the CH$_4$ flow rates
are 50 sccm, 30 sccm, 20 sccm, 18 sccm and 15 sccm), thus grown
with varying H$_2$ to CH$_4$ ratio during deposition, will be
represented as Sample A to Sample E, in the rest of the article.

Raman spectra are measured in back-scattering geometry using a 488
nm Argon ion laser as an excitation source. The spectrometer is
equipped with 1200 g$/$mm holographic grating, a holographic
super-notch filter, and a Peltier cooled CCD detector. With 100
$\mu$m slit-width of the spectrometer the resolution of our Raman
measurement is 1 cm$^{-1}$.

XP spectra are taken using PHI-5702 X-ray Photoelectron
Spectroscope  operating with monochromated Al K$\alpha$
irradiation (photon energy 1476.6 eV) as an excitation source at a
pass energy of 29.4 eV. The chamber pressure is maintained at
10$^{-8}$ Pa.

Ellipsometric measurements of the five DLC films (Sample A -
Sample E) are performed for the spectral range from 370 to 990 nm
(1.25–3.35 eV) using J.A. Wollam ellipsometer( make USA) in the
reflection mode. The angle of incidence is 70$^\circ$. From the
ellipsometric measurements, the thickness of the films are
estimated to be $\sim$ 120 $\pm$ 20 nm.

Hardness of the films are measured by Leica micro- hardness tester
operated using Vickers diamond indenter by applying a load of 50
gf at a minimum of  three places for each sample. Diagonals of the
indentations are measured to eliminate the asymmetry of the
diamond pyramid.

Samples for Transmission Electron Microscopy (TEM) are deposited
on 400 mesh copper TEM grids coated with carbon films.  The
suspended film is obtained by  sonicating the substrate with film
in Acetone; which is then directly added drop-wise on the grid.
The excess acetone is allowed to evaporate in air. The grids are
examined in Hitachi H600 microscope operated at 75 kV.

\section{Characterization of DLC films by Raman measurements and by other techniques}

\begin{figure}
\centerline{\epsfxsize=6.5in\epsffile{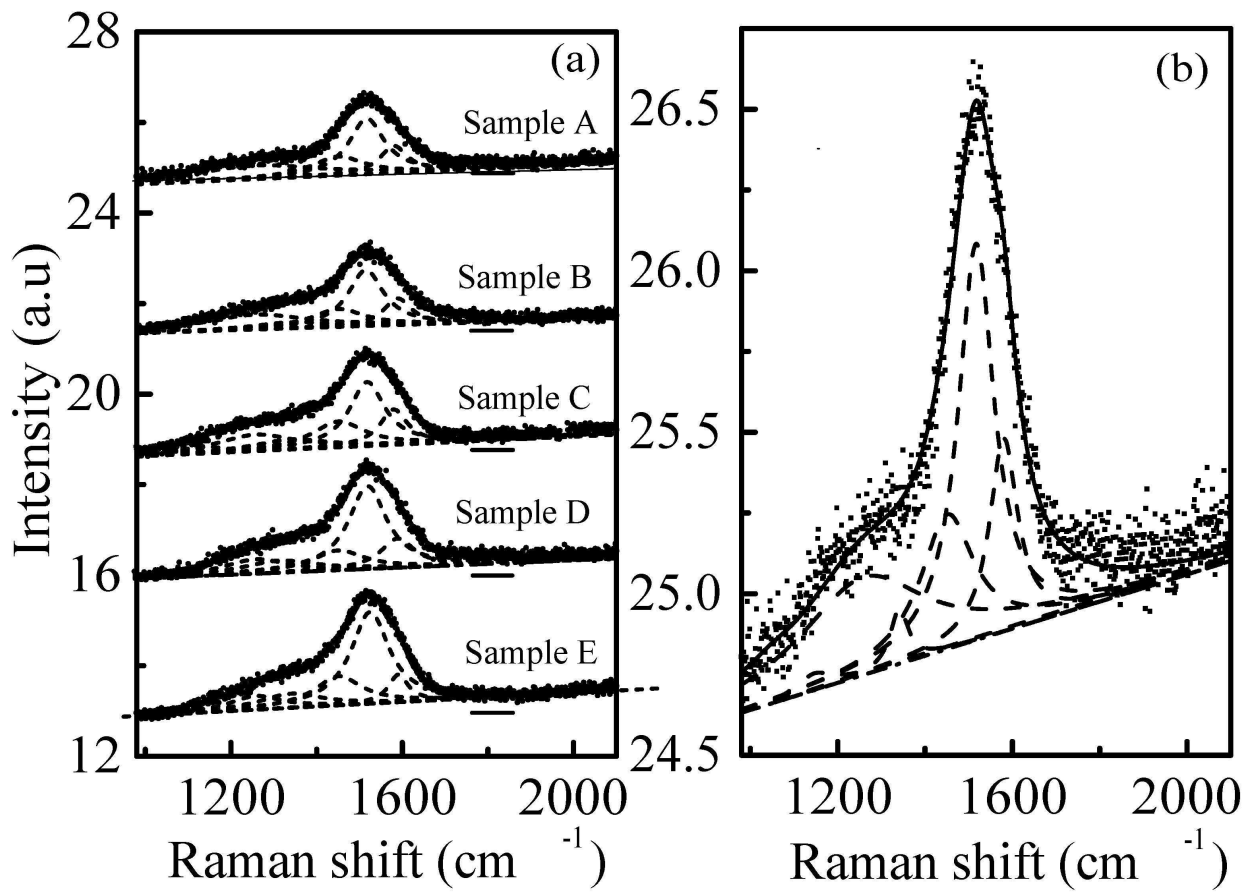}} \caption{(a)
Raman spectra for Sample A to Sample E. The filled circles are
experimental data points. For each spectrum, the linear background
is shown by the dashed-dotted line, the deconvoluted components
are shown by dashed lines and the resultant fitted curve is shown
by the solid line. Horizontal line on the right of each spectrum
shows the zero intensity scale, (b) Raman spectrum for Sample A
and the fitted lines.}
\end{figure}

Fig. 1 (a) shows Raman spectra of Sample A to Sample E for the
range from 950 to 2500 cm$^{-1}$. The increasing PL background of
each spectra is a typical signature of hydrogenated DLC (HDLC)
films \cite{ferrari:2000,marchon:1997}. Keeping in mind the
possibility of the presence of different phases of carbon,
mentioned in section I, we have deconvoluted each spectrum with
six Lorentzian functions (shown by dashed lines in Fig. 1) keeping
intensities, width and peak positions as free fitting parameters.
Table I provides all parameters, which we have obtained, in this
analysis. It is to be noted that the asymmetric Breit-Wigner-Fano
(commonly known as BWF) line shape is usually used to fit the
G-peak which appears due to asymmetry of the vibrational density
of states of graphite towards lower wavenumber. The tail of BWF
line  takes into account the Raman modes at 1100 and 1400
cm$^{-1}$, mentioned in the previous section, without giving rise
to extra peaks. In Fig. 1 (and Table I) we note that the
intensities of the D-peak and other low frequency peaks are quite
less than the intensity of the G-peak. Thus, if we deconvolute
each spectrum with all low frequency components to obtain the
information of individual modes, there is an inherent difficulty
in  using BWF line shape for the G-peak. For this reason, we have
used Lorentzian line shapes for all modes to analyze our spectra.
For clarity, in Fig 1 (b) we show the experimental data,
deconvoluted lines and net fitted spectrum for Sample A.
Below we discuss the properties of our DLC films obtained from
above Raman data analysis :

\begin{table}[htbp]
\caption{Assignment of vibrational bands for DLC Raman spectra }
\begin{tabular}{|c|c|c|c|c|c|c|} \hline
Sample  & Sample E & Sample D & Sample C & Sample B & Sample A & Remarks \\
\hline
Peak position (cm$^{-1})$ & 1591  &  1589 &   1590 &    1589   & 1590 & Dumbbell \\

Width (cm$^{-1}$) &   38   &  45   & 39 &  39 &  39 & defect \\

Intensity   &    1708   &  3082   &  3045  &  2283  &  2331 & in NC\\
\hline
Peak position (cm$^{-1})$ &1539  &  1524  &  1518  &  1515  &  1517 & \\

Width (cm$^{-1}$) &   54 & 54 & 50 & 50 & 50 & G-peak \\

Intensity   &  10943 &  9885  &  7059  &  6182  &  6132 & \\
\hline
Peak position (cm$^{-1})$ &1451  &  1451  &  1451  &  1451  &  1451& \\

Width (cm$^{-1}$) &  67  & 64 & 66 & 64 & 65 & Trans-PA \\

Intensity   &  4304   &  2919 &   4225 &   2919  &  900 & ($\nu_3$ mode)\\
\hline
Peak position (cm$^{-1})$ & 1341  &  1341  &  1341  &  1341  &  1341 & \\

Width (cm$^{-1}$) &  90 & 105 &105 & 105  & 105 & D-peak \\

Intensity   &  2813  &  2800  &  2332  &  1416  &  2146 & \\
\hline
Peak position (cm$^{-1})$ & 1258  &  1258 &   1258  &  1258 &   1258 & \\

Width (cm$^{-1}$) &  73 & 91 & 110 &148 &177 & NC\\

Intensity   &  2214  &  2894 &   4000  &  4858  &  5363  & diamond\\
\hline
Peak position (cm$^{-1})$ & 1150  &  1146  &  1146  &  1146 &   1146  & \\

Width (cm$^{-1}$) &  51 & 50 & 52 & 50 & 50& Trans-PA \\

Intensity   & 84 & 70 & 73 & 58 & 74 & ($\nu_1$ mode)\\
\hline
\end{tabular}
\end{table}

\begin{figure}[htbp]
\centerline{\epsfxsize=3.5in\epsffile{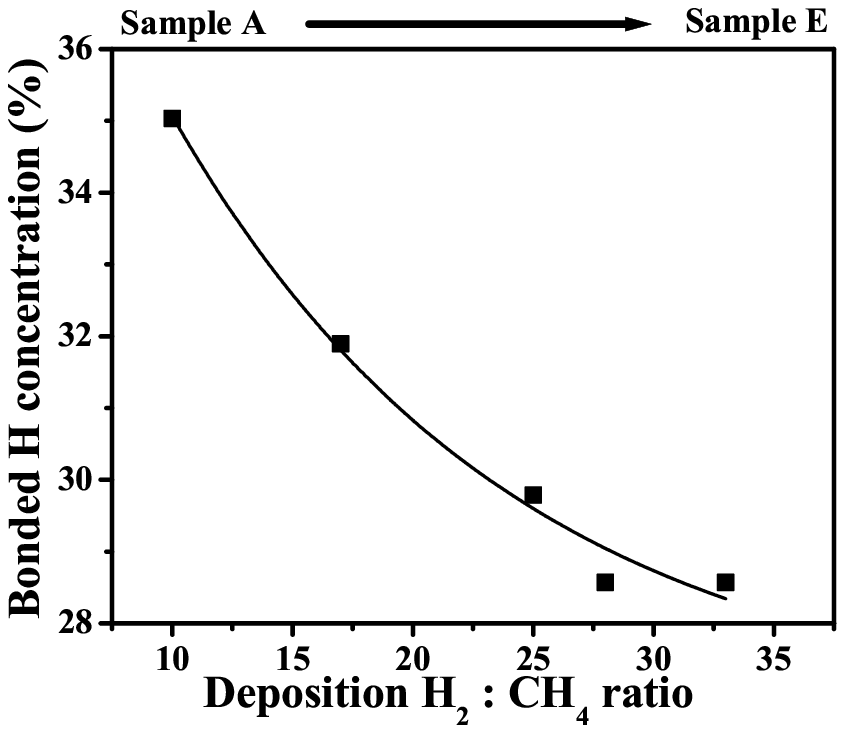}}
 \caption{The
change in hydrogen content from Sample A to Sample E.}
\end{figure}

(a) \emph{Hydrogen content}: The main effect of hydrogen in HDLC
films is to modify its C-C network. Instead of increasing the
fraction of C-C bond, hydrogen saturates the C=C bonds as $\equiv$
CH$_x$ groups to increase the sp$^3$ content in the film
\cite{ferrari:2000}. In visible Raman spectra, C-H stretching mode
lies above 3000 cm$^{-1}$, whereas, C-H bending mode (1290-1400
cm$^{-1}$) is masked by the D-peak region \cite{ristein:1998}.
Moreover, these modes are not resonantly enhanced in visible Raman
spectroscopy. Thus, it is difficult to estimate hydrogen content
in a HDLC film directly by using this technique. However, as a
result of the recombination of electron-hole pairs within sp$^2$
bonded clusters in HDLC films, the hydrogen content in films gives
rise to a strong PL background for first order Raman spectra
\cite{marchon:1997}. The ratio between the slope $m$ of the fitted
linear background of the Raman spectrum (shown by dashed-dotted
line in Fig. 1 (a)) due to photoluminescence and the intensity of
G-peak (I$_G$), $m/I_G$, can be used as a measure of the bonded H
content in the film. The slope parameter is described in
micrometer unit \cite{marchon:1997}. Using this analysis, the
decrease in intensity of bonded hydrogen content from Sample A to
Sample E is shown in Fig. 2.
It is to be noted that the observed decrease in H$_2$ content is
expected to increase the sp$^2$ cluster size and results in a
decrease in band gap from Sample A to Sample E
\cite{ferrari:2000}.

\begin{figure}[htbp]
\centerline{\epsfxsize=6.5in\epsffile{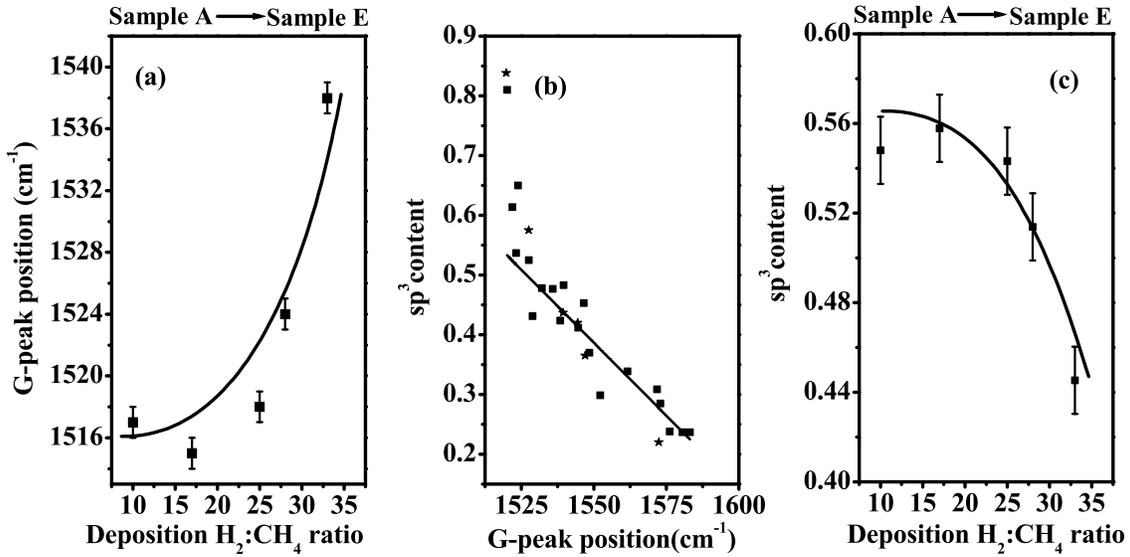}}
\caption{Variation in (a) sp$^3$ content in the a-C:H DLC films
with G-peak positions as obtained from Ref. 7 (filled squares) and
Ref. 10 (filled stars) (the fitted curve to these data points
using Eqn. 1 is shown by solid line) (b) G-peak position from
Sample A to Sample E  (c) sp$^3$ content in the films from Sample
A to Sample E. In (b) and (c) the solid lines are guides to the
eyes.}
\end{figure}

(b) \emph{sp$^3$ fractions in the films }: The variation in
$\omega_G$ from Sample A to Sample E, as obtained from their Raman
spectra in Fig.1 (a), is shown in Fig. 3(a). The frequency of the
Raman shift for $\omega_G$ and the presence of NC diamond in the
films (which will be discussed latter), indicate that the
characteristics of the films fall under the second stage of the
three stage model. If one carefully notes the dispersion in Raman
G-peak position, $\omega_G$, in amorphous HDLC films and the
sp$^3$ content in the same samples obtained by other measurements
(NMR and EELS), a certain correlation between these two parameters
can be clearly observed. Ferrari and Robertson in Fig. 14 of ref
\cite{ferrari:2000} summarized experimental data from different
sources and have shown that sp$^3$ content in the HDLC film is
related to $\omega_G$ in the films. Similar characteristic has
also been shown in Ref. \cite{tamor:1991} by Tamor et al. We have
taken average of all data presented in these references in Fig.
3(b) and obtained an empirical relation by fitting the data points
between 1525 cm$^{-1}$ and 1580 cm$^{-1}$ by a polynomial equation

\begin{equation}
\mbox {sp}^{3} \mbox{content}= 0.24 - 48.9 (\omega_{G} - 0.1580)
\end{equation}

In this equation, $\omega_G$  has been taken in unit of inverse of
micrometer unit. Eqn. 1 also demonstrates that for $\omega_G$ at
1580 cm$^{-1}$, the sp$^3$ fraction in the film is $\sim$ 0.24.
Using the shift in G peak position in the above equation, the
changes in sp$^3$ content in the films are shown in Fig. 3(c)
(they are also tabulated in Table II). If one looks into the
ternary phase diagram for HDLC film \cite{robertson:1991}, as
shown in Fig. 4, it is clear that with a decrease in hydrogen
content the C-C sp$^3$ bonding in the film should also decrease.
Thus, the observed decrease in hydrogen content (shown in Fig. 2)
supports the decrease in  sp$^3$ content in films, shown in Fig.
3(c).

\begin{figure}[htbp]
\centerline{\epsfxsize=3.0in\epsffile{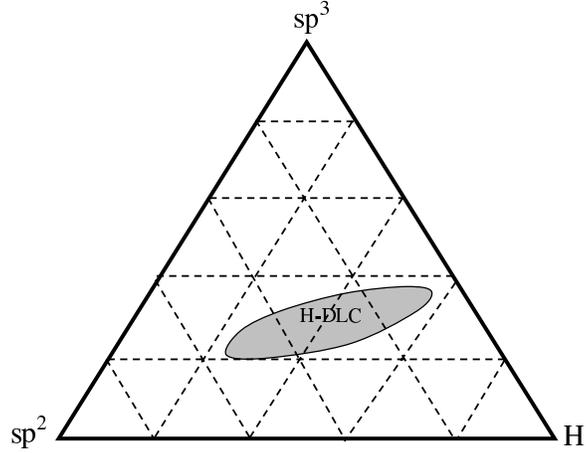}} \caption{The
ternary phase diagram for a-DLC films (from Ref.
\cite{ferrari:2000}).}
\end{figure}

\begin{figure}[htbp]
\centerline{\epsfxsize=4.5in\epsffile{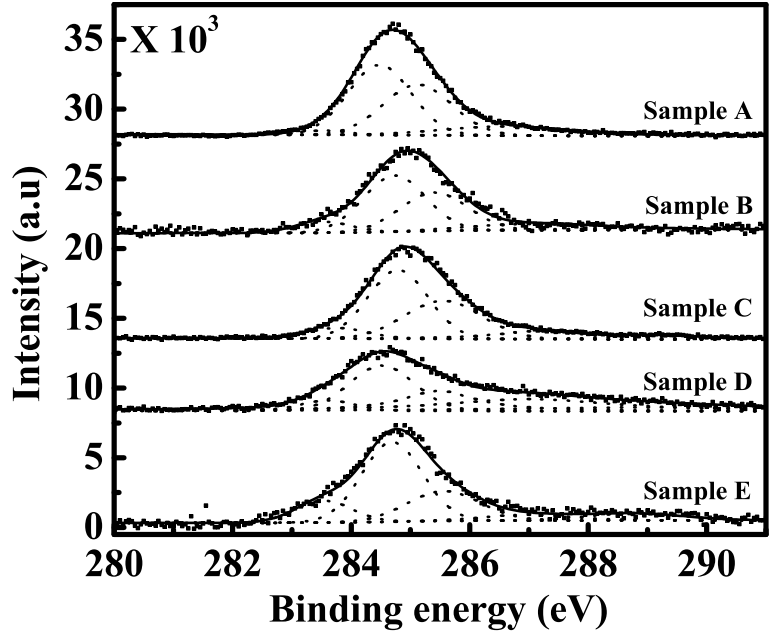}}
 \caption{XP
spectra for Sample A to Sample E. The filled circles are
experimental data points. For each spectrum, the deconvoluted
components are shown by dotted lines and the resultant fitted
curve is shown by the solid line.}
\end{figure}

Next, we have used  XP Spectroscopy on these samples to measure
their sp$^3$ content. The XP spectra for Sample A to Sample E are
shown in Fig. 5. After subtracting the background signal, C1s
spectra of HDLC films are deconvoluted into four components around
283.6 $\pm$0.2 eV, 284.6 $\pm$0.2 eV, 285.4 $\pm$0.2 eV, and 286.9
$\pm$0.3 eV  (Fig. 5). The peaks at around 284.3 eV and 285.3 eV
correspond to sp$^2$ carbon atoms \cite{cardona:1978} and sp$^3$
C–C bond \cite{zhao:2000}, respectively. The feature at 286.6 eV
is assigned to C–O contamination formed on the film surface due to
air exposure \cite{yan:2004,zhao:2000}. The presence of lower
energy peak at around 283 eV indicates that some carbon atoms in
HDLC films are bonded to silicon substrate (carbon peak of
carbide) \cite{yan:2004}. The full width at half maxima for both
peaks at 284.3 eV and 285.3 eV are kept at 1.4$\pm$0.05 eV. The
sp$^3$ content in the films, as obtained from the ratio of the
corresponding sp$^3$ peak area over the total C1s peak area, have
been tabulated in Table II along with the same obtained from
visible Raman data analysis.

\begin{table}[htbp]
\caption{Comparison of sp$^3$ content in  HDLC films as obtained
from Raman and XP spectroscopic measurements.}
\begin{tabular}{|c|c|c|}\hline
Sample & \multicolumn{2}{c|}{sp$^3$ content in $\%$}  \\\hline
       & from Raman & from XPS \\ \hline
Sample A & 55 & 44\\ \hline

Sample B & 56 & 43\\ \hline

Sample C & 54 & 42\\ \hline

Sample D & 51 & 43\\ \hline

Sample E & 44 & 31\\ \hline
\end{tabular}
\end{table}

\begin{figure}[htbp]
\centerline{\epsfxsize=5.5in\epsffile{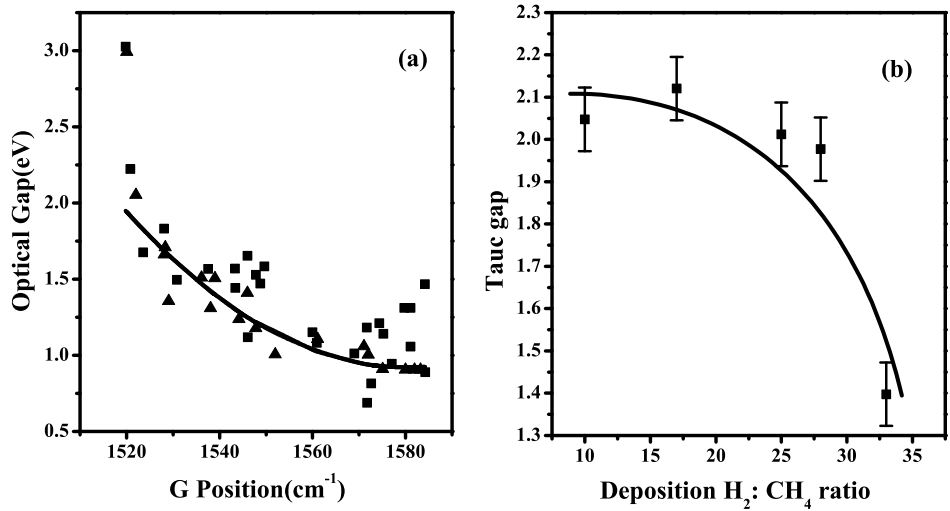}}
\caption{Variation in (a) optical gap in the a-C:H DLC films with
G-peak positions as obtained from Ref. 7 (filled squares) and Ref.
6 (filled triangles)( the fitted curve to these data points using
Eqn. 2 is shown by solid line)(b) Tauc gap from Sample A to Sample
E using Eqn. 2}
\end{figure}

(c)\emph{Tauc gap} :  For quantitative analysis of the Tauc gap in
DLC films, as done in reference \cite{ferrari:2000}, the authors
have shown a correlation between $\omega_G$ with the optical Tauc
gap of the material. Similar correlation between the same
parameters has been reported by Tamor et al in Ref.
\cite{tamor:1994}. We make a summary of all experimental data from
the above references in Fig. 6(a) and fit them with a polynomial
equation

\begin{equation}
\mbox{Tauc Gap} = 0.92 + 28410 (\omega_{G} - 0.1580)^2
\end{equation}

As in Eqn. 1, $\omega_G$  has been taken in unit of inverse of
micrometer unit. Eqn. 2 also demonstrates that for $\omega_G$ at
1580 cm$^{-1}$, the Tauc gap of the film is $\sim$ 0.92 eV. From
the variation in $\omega_G$ as shown in Fig. 3(b) and using Eqn.
2, we have estimated the variation in Tauc gap of the samples,
which has been shown in Fig. 6(b) and also listed in Table III.
Here, we would like to recall that a decrease in Tauc gap was
expected due to the decreasing hydrogen content from Sample A to
Sample E (see Fig. 2).

\begin{table}[htbp]
\caption{Comparison of Tauc gap of the HDLC films as obtained from
Raman and Ellipsometric  measurements.}
\begin{tabular}{|c|c|c|}\hline
Sample & \multicolumn{2}{c|}{Tauc Gap (eV)}  \\\hline
       & from Raman & from Ellipsometry \\ \hline
Sample A & 2.05 & 1.59\\ \hline

Sample B & 2.12 & 1.53\\ \hline

Sample C & 2.01 & 1.49\\ \hline

Sample D & 1.98 & 1.48\\ \hline

Sample E & 1.40 & 1.41\\ \hline
\end{tabular}
\end{table}

\begin{figure}[htbp]
\centerline{\epsfxsize=3.5in\epsffile{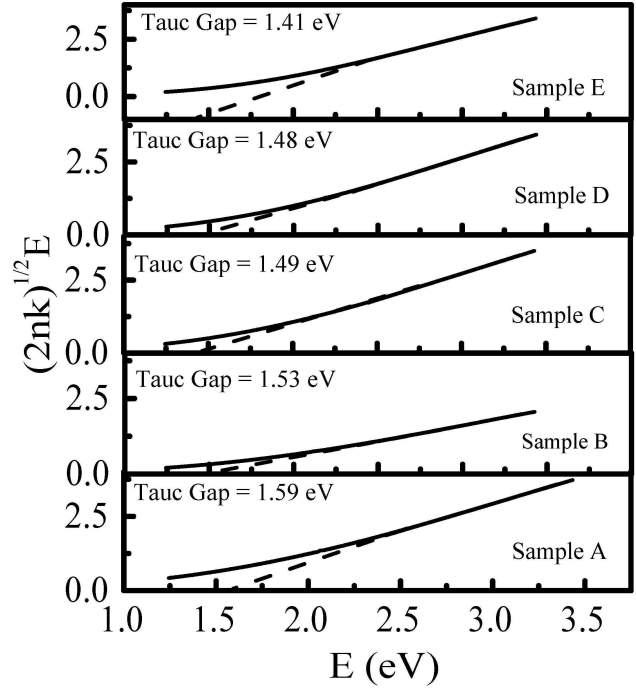}} \caption{Tauc
plots for Sample A to Sample E: Solid and dashed line denote the
experimental data obtained from ellipsometric measurements and
linear fit to the data points, respectively.}
\end{figure}

The optical  gap obtained from the analysis of Raman data, has
been compared with the same measured directly from Ellipsometry.
Both layer and substrate are characterized by spectral
dependencies of the refractive indices (n and n$^\star$) and the
extinction coefficients (k and k$^\star$).The spectral dependency
of n$^\star$ and k$^\star$ of the silicon substrate are taken from
ref. \cite{ohlidal:1999}. The spectral dependencies of the optical
constants (n and k) of the DLC films for a wide spectral range can
be interpreted by a semi-classical dispersion model with Lorentz
oscillators corresponding to $\pi \rightarrow \pi*$ (originating
from sp$^2$ sites) and $\sigma \rightarrow \sigma*$ (originating
from sp$^3$ sites) interband transitions \cite{zhang:1998}. We
have used this model to obtain n and k from the ellipsometric
data. The shape of the imaginary part of the complex dielectric
function $\zeta [\varepsilon(E)]$ ( = 2nk) above the band gap is
given by the Tauc equation
\begin{equation}
\zeta [\varepsilon(E)]=\frac{\alpha (E-E_{g})^2}{E^2}
\end{equation}
$\alpha$ is the absorbance of the film. The intersection of the
plot $\sqrt{\zeta[\varepsilon(E)]}E$ vs. $E$ (Tauc plot) with the
$E$ axis, measures the energy of the band gap, $E_g$.  Tauc plots
for Sample A to Sample E as obtained from the ellipsometric
measurements are shown in Fig. 7. It is to be noted that  plots
for our DLC films are not strictly straight lines as indicated by
Eqn. 3. This can be due to the existence of the Urbach tail (the
transitions between localized states inside the band gap and
extended states inside valence or conduction bands, which  causes
an exponential broadening of the absorption edge). From the
intercept of the tangent of Tauc plots on $E$ axis we have
obtained the optical band gap of HDLC films. The variation in
optical band gap from Sample A to Sample E, as observed by
ellipsometric measurements has been compared with that obtained
from Raman measurements in Table IV.


(d)\emph{Hardness}: In reference \cite{marchon:1997} we find a
comment on variation in hardness of the DLC films with the
percentage of Hydrogen in the film. Using this reference, we
propose an empirical relation by fitting the data points by a
linear equation

\begin{equation}
\mbox {Hardness in GPa}= 44.195 -0.93 \times \mbox {(\% of
hydrogen content)}
\end{equation}

Using Eqn. 4 and the variation in hydrogen contents in our
samples, we have estimated the hardness of the films. The hardness
of each film, as obtained indirectly from Raman data analysis has
been tabulated in Table IV along with the same obtained from
micro-hardness testing.

\begin{table}[htbp]
\caption{Comparison of hardness of HDLC films as obtained from
Raman and micro-hardness  measurements.}
\begin{tabular}{|c|c|c|}\hline
Sample & \multicolumn{2}{c|}{Hardness (GPa)}  \\\hline
       & from Raman & from micro-hardness test \\ \hline
Sample A & 11.35 & 13.1 \\ \hline

Sample B &15.37 & 12.7\\ \hline

Sample C & 17.22 & 13.2\\ \hline

Sample D & 17.60 & 11.4\\ \hline

Sample E & 18.15 & 13.3 \\ \hline
\end{tabular}
\end{table}

From Fig. 2, Fig. 3(c) and the above hardness analysis we conclude
(a) a decrease in sp$^3$ bonding from Sample A to Sample E does
not degrade the hardness and scratch resistance of the deposit and
(b) the low hydrogen content in the deposit strengthens the
inter-molecular structure and, hence, leads to improved mechanical
properties. Similar observation has been reported for Si implanted
DLC films, where it has been shown from XP spectroscopic
measurements that low hydrogen content in the DLC films results in
a better hardness in these films \cite{zhao:2000}.

\begin{figure}[htbp]
\centerline{\epsfxsize=5.5in\epsffile{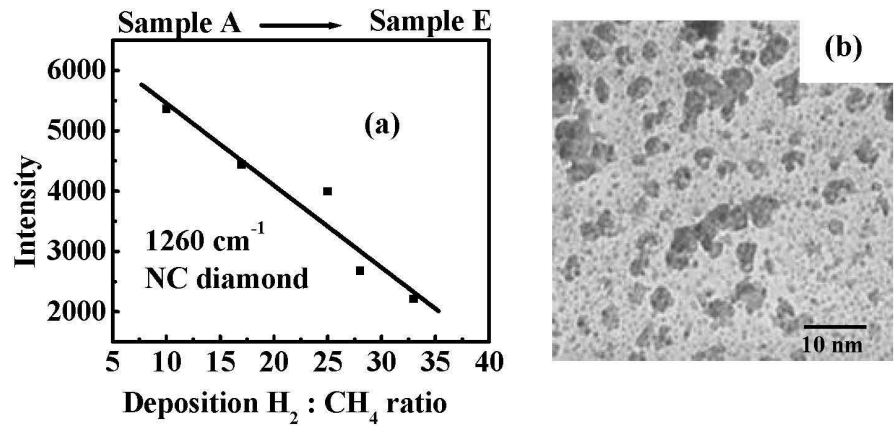}}
\caption{Variation in intensity of  the peak at 1260 cm$^{-1}$ due
to NC diamond.}
\end{figure}

(e) \emph{Nanocrystalline diamond} : The decrease in intensity of
the peak at 1260 cm$^{-1}$  [shown in Fig. 8 (a)] indicates a
decrease in NC diamond phase from Sample A to Sample E. From our
above analysis we summarize that the  NC component in these films
decreases along with (i) an increase in G-peak position from 1517
cm$^{-1}$ to 1539 cm$^{-1}$ [Fig. 3(b)] and (ii) a decrease in
sp$^3$ content [Fig. 3(c)]. The TEM image of one of our films is
shown in Fig. 8(b). Presence of NC diamond along with larger
particles is clear in the micrograph.  This, in a way, supports
our Raman data analysis by deconvolution.

\section{Discussion}
In this article we have tried to establish a workable
phenomenological picture for characterizing DLC films using
visible Raman spectroscopy. We are able to estimate sp$^3$ and
hydrogen content, optical gap and hardness of films, prepared by
plasma deposition technique, by visible Raman measurements. We
have summarized a number of data, available in the literature and
then utilized the `\emph{three stage model}', which critically
assesses the position of G peak in the Raman spectra of the films,
to propose empirical relations between G-peak position and the
values of sp$^3$ content and Tauc gap in HDLC films.

We have also measured the sp$^3$ content of the film by XP
spectroscopy and optical band gap by ellipsometry. In addition,
the hardness of the films are estimated from micro-hardness
testing. In Table II, III and IV, we have compared the
characteristics of DLC films as obtained from Raman data analysis
and directly by the above mentioned measurements.

We now discuss the limitations of techniques while analyzing the
characteristics of HDLC films on Si substrate. XP spectroscopy is
a surface sensitive probe. Thus, this technique underestimates the
value of sp$^3$ fraction (obtained by fitting the XP spectra) if
the thickness of the film is more than the electron mean free path
($\sim$ 50 \AA). Indeed, in Table II, we find that though both
Raman and XP spectroscopic analysis exhibit decrease in sp$^3$
content from Sample A to Sample E, the absolute values of the same
obtained from XP spectroscopic measurements are less than the
values obtained from Raman data analysis. Here, we would like to
point out that rich sp$^3$ phase of DLC films is often located
below (depends on C energy) the surface layer. The surface layers
are much more graphitic \cite{Lifshitz:1999}. XP spectroscopic
measurement after etching the surface of the  film  under ultra
high vacuum can be a solution to get the information of the inner
layers; however, this experiment will damage the sample.

Furthermore, ellipsometry does not measure optical constants of
materials directly. In this technique, we measure spectral
dependency of the complex quantity $\rho$, which is defined as
$\rho = \frac{r_p}{r_s}$. $r_s$ and $r_p$ represent the Fresnel
coefficients of the sample. To get the spectral dependencies of
the optical constants in the entire spectral region, one needs to
understand the nature of dispersion in HDLC films. We have used
`single sample method' to analyze the ellipsometric data. We
interpreted the spectral dependencies of the  optical constants
using Lorentz oscillator model. However, this model does not take
into account the hydrogenated amorphous phase of the HDLC layer in
a true sense. A more detailed nontrivial numerical analysis on
modified Lorentz oscillator model is required to obtain the
correct band gap of DLC films using ellipsometric measurements
\cite{franta:2002}.  may explain the mismatch in the Tauc gap of
the samples obtained from Raman and ellipsometry measurements
(shown in Table III); though both exhibit decrease in Tauc gap
from Sample A to Sample E

Regarding the hardness measurements the source of discrepancy lies
in the fact that our films are grown on Si substrate. The hardness
of crystalline Si is $\sim$ 13 GPa, which is of the same order of
magnitude as that of DLC. Thus, by indentation it is difficult to
obtain the correct hardness of HDLC films subtracting the effect
of the substrate. Indentation with less load does not measure the
hardness of the films correctly. We attribute the discrepancy in
hardness of the films, obtained from Raman and micro-hardness test
to the fact that the latter is not a fool-proof technique to
obtain the hardness of relatively thin DLC films on Si substrate.

Our Raman data analysis, based on a few empirical relations, may
not be very accurate. However, by this, we overcome some of the
above-mentioned shortcomings of the other techniques. The skin
depth of the 488 nm excitation source, which we have used in Raman
measurements is $\sim$ 6 $\mu$. Thus, using this technique we get
information for the whole material rather than only for the
surface as was the case in XP spectroscopic measurements. Unlike
ellipsometric measurements, Raman analysis takes into account all
phases of carbon. The hardness of the films, obtained from Raman
data analysis, does not require an understanding of the effect of
the substrate separately. By correct data analysis, we have also
shown the presence of NC diamond in our films.

In addition to  $\omega_G$, the intensity ratio of the D- and G
peaks, can also be used to estimate the above parameters of HDLC
films.  We did not follow this procedure because of the following
reasons : i) the intensity of the D peak is quite low compared to
that of G peak in our samples (so the numerical value of
I$_D$/I$_{G}$ can be erroneous) and ii) in the low frequency
regime of the Raman spectra the weak D-peak coexists with Raman
components of other phases of carbon; hence the measured intensity
of the D-peak from the curve fitting is not unique, it varies
strongly with the choice of width and intensity of other nearby
peaks.

Table V summarizes the physical properties of our HDLC films, as
obtained by Raman data analysis.

\begin{table}[htbp]
\caption{Properties of HDLC films as obtained from Raman
measurements}
\begin{tabular}{|c|c|c|c|}\hline
Sample & $\%$ of sp$^3$ content & Tauc gap (eV) & hardness (GPa)\\
\hline

Sample A & 57 & 2.03 &11.35\\ \hline

Sample B & 56  & 2.11  &15.37\\ \hline

Sample C & 55 & 1.91  &17.22\\ \hline

Sample D & 54 & 1.96  &17.60\\ \hline

Sample E & 43 & 1.39 & 18.15 \\ \hline
\end{tabular}
\end{table}

\section{Acknowledgement}
AR thanks Department of Science and Technology, India, for
financial assistance. Authors thank Dr. S. Varma, Institute of
Physics, Bhubaneswar, for their help on XPS measurements, Prof.
K.K. Ray, Indian Institute of Technology, Kharagapur for Micro
-Vicker's testing, Mr. P. Roy, Saha Institute of Nuclear Physics
for TEM measurements and Dr. Jens Raacke, University of Wuppertal,
Germany for Ellipsometric measurements.

\end{document}